\documentclass[12pt,groupcitations]{article}
\pdfoutput=1
\usepackage[T1]{fontenc}
\usepackage[ansinew]{inputenc}
\usepackage[italian,english]{babel}
\usepackage{amsfonts}
\usepackage{amsmath}
\usepackage{array}
\usepackage{amsthm}
\usepackage{amssymb}
\pdfoutput=1
\usepackage{graphicx}
\usepackage{subfigure}
\usepackage{braket}
\usepackage{eucal}
\usepackage{verbatim}
\usepackage{caption}
\usepackage{multicol}
\usepackage{tikz}

\usepackage{float}

\usetikzlibrary{arrows,shapes}
\usetikzlibrary{matrix,arrows}
\newcommand{\be}{\begin{eqnarray}}
\newcommand{\ee}{\end{eqnarray}}

\renewcommand{\d}{\mbox{${\rm d}$}} 
\newcommand{\lp}{\ell_{\rm p}}
\newcommand{\mpl}{M_{\rm Pl}}

\newcommand{\rh}{r_{\rm H}}

\newcommand{\Mpl}{M_{\rm Pl}}

\begin{document}
\begin{center}
{\Large \bf Horizon Quantum Mechanics of Generalized Uncertainty Principle Black Holes}

\bigskip\bigskip

\large{Luciano Manfredi\footnote{\textit{E-mail:} \texttt{lmanfred@lion.lmu.edu}} and Jonas Mureika\footnote{\textit{E-mail:} \texttt{jmureika@lmu.edu}}\\
\bigskip
 \textit{Department of Physics, Loyola Marymount University, Los Angeles, CA  90045-2659}}
 \end{center}
 
 \vskip 1.5cm

\begin{abstract}
We study the Horizon Wavefunction (HWF) description of a generalized uncertainty principle inspired metric that admits sub-Planckian black holes, where the black hole mass $m$ is replaced by $M = m\left(1+\frac{\beta}{2}\frac{\mpl^2}{m^2}\right)$.    Considering the case of a wave-packet shaped by a Gaussian distribution, we compute the HWF and the probability ${\cal {P}}_{BH}$ that the source is a (quantum) black hole, {\it i.e.} that it lies within its horizon radius.  The case $\beta<0$ is qualitatively similar to the standard Schwarzschild case, and the general shape of ${\cal {P}}_{BH}$ is maintained when decreasing the free parameter, but shifted to reduce the probability for the particle to be a black hole accordingly. The probability grows with increasing mass slowly for more negative $\beta$, and drops to 0 for a minimum mass value. The scenario differs in significantly for increasing $\beta>0$, where a minimum in ${\cal {P}}_{BH}$ is encountered, thus meaning that every particle has some probability of decaying to a black hole. Furthermore, for sufficiently large $\beta$ we find that every particle is a quantum black hole, in agreement with the intuitive effect of increasing $\beta$, which creates larger $M$ and $R_{H}$ terms. This is likely due to a ``dimensional reduction'' feature of the model, where the black hole characteristics for sub-Planckian black holes mimic those in $(1+1)$-dimensions and the horizon size grows as $R_H \sim M^{-1}$.  
\end{abstract}

\section{Introduction}
Black holes are special objects in gravitational physics because they are expected to reveal features of both classical and quantum gravitation.  Indeed, one can note this connection in the simple fact that the defining parameter of the quantum gravity scale, the Planck mass $\Mpl$, simultaneously sets the strength of classical gravitation $G = \Mpl^{-2}$.  A complete understanding of black hole physics will thus help shed light on this elusive theory.  Although large black holes may reveal hints of quantum effects through {\it e.g.} the morphology of their shadows \cite{Johannsen:2015hib}, it is anticipated that eventual observation of quantum scale black holes formed in high energy collisions will provide direct evidence.  In this regime, these objects transcend the classical and quantum, and thus forming reliable predictions of their physics becomes tenuous in the absence of a complete theory of quantum gravity.

Although such a formulation is still incomplete, the literature is replete with first steps beyond the classical regime and into the quantum realm.  Such semiclassical approaches generally rely on a classical framework, extended to include quantum effects at the appropriate energy or length scale that tame or remove the singularity.   Examples include noncommutative geometry inspired models (\cite{Nicolini:2005vd}; also see \cite{Nicolini:2008aj} for an overview and additional references therein),  the Generalized Uncertainty
Principle \cite{Carr:2015nqa}, and asymptotic safety \cite{Reuter:2005bb}.  Other approaches stem from quantum mechanical first principles and  introduce gravitation as an energy or potential constraint.  The Schrodinger-Newton equation \cite{Ruffini:1969qy,Bahrami:2014gwa} can be derived from the weak-field Einstein's equations with the stress energy tensor replaced by the expectation of a quantum operator, from which potential table-top quantum gravity mesaurements may be possible \cite{Grossardt:2015moa}.  General aspects of quantum field theories in curved spacetime are also well-known modifications of quantum theories to include gravitation \cite{Davies:1976pm,Birrell:1982ix,Jacobson:2003vx}.

A more recent approach in understanding the nature of quantum black holes is to consider the quantum mechanical conditions for their creation in terms of a wavefunction description.  Dubbed the ``Horizon Wavefunction'' approach (HWF), the black hole is treated as a quantum particle whose spatial wavefunction is contained within its classical horizon radius \cite{Casadio:2013tma,Casadio:2015qaq}.  If such particles are created in high energy collisions, then the chance of creating a black hole can be assessed by evaluating the associated probability.  In particularly, the HWF has been used to understand aspects of quantum black hole thermodynamics, including evaporation signatures in four-dimensional spacetime \cite{Casadio:2015lis,Calmet:2015pea,Casadio:2013tza,Casadio:2013aua,Casadio:2013uga}, as well as extra- and $(1+1)$-dimensional scenarios \cite{Casadio:2015jha}, 
Potential experimentally detectable signatures that arise from such a description have been discused in \cite{Arsene:2016kvf}.

A particularly interesting feature of the HWF formalism is the appearance of a generalized uncertainty principle (GUP), in which quantum uncertainties are simultaneously influenced by the wave and gravitational length scales of a particle. Originally noted as a feature of string theory \cite{Veneziano:1986zf}, the GUP has been shown to be a largely model-independent prediction of quantum gravity theories, including loop quantum gravity~\cite{ashtekar}, non-commutative quantum mechanics \cite{majid}, gravity ultraviolet self-completeness \cite{nicolini} and other minimum length scenarios \cite{maggiore_1,maggiore_2,maggiore_3}.   While most minimal length approaches yield a lower bound to black hole masses, a recently-derived GUP modified Schwarzschild metric was shown to allow the existence of sub-Planckian black holes \cite{Carr:2015nqa}.   A special feature of these sub-Planckian black holes is that their physical and thermodynamic characteristics mimic those of $(1+1)$-dimensional black holes -- {\it i.e.} the horizon size varies inversely with the black hole mass, $R_H \sim M_{\rm BH}^{-1}$, and the Hawking temperature linear in the mass, $T_H \sim M$.  

Since the HWF can both predict the probability of black hole formation for arbitrary masses and also source constraints on a GUP from the quantum mechanical side, we seek to understand how encoding the GUP in the metric will influence the probability of black hole formation.  In the following paper, we apply the HWF prescription to the GUP-inspired metric of \cite{Carr:2015nqa}.  After first reviewing the formalism for both HWF and the GUP metric in Sections~\ref{review} and ~\ref{gupreview},  we derive expressions for the HWF and black hole probabilities in both the super- and sub-Planckian mass regimes for varying GUP model parameters.   In the former case, we find the results to be in agreement with those of the Schwarzschild HWF.   In the sub-Planckian regime, we show that the probability of a particle of arbitrarily small mass becoming a black hole tends to unity.   We discuss the results in the conclusions.

\section{The Horizon Wavefunction Formalism}
\label{review}
In order to follow the prescription outlined above, we review the first ingredient of the approach -- namely, the HWF framework.  The following arguments reproduce the standard approach detailed in {\it e.g.} \cite{Casadio:2013tma} and similar references.  We start from the definition of the trapping surface,
\be
g^{ij}\nabla_{i}r\nabla_{j}r=0
\ee
where $\nabla_{i}r$ is normal to spherical surfaces of area ${\cal A}=4\pi r^{2}$. From this, one can derive the metric function
\be
g^{rr}=1-\frac{2\lp(m/\mpl)}{r} \;,
\ee
if one assign coordinates $(x^1,x^2)=(t,r)$.  The quantities $\Mpl$ and $\lp$ are the Planck mass and length, respectively.  Assuming a rough flat space, the Misner-Sharp mass can be calculated as
\be 
m(r,t)=4\pi \int_{0}^{r}\rho(\bar{r},t)\bar{r}^2\d \bar{r}~~,
\ee
where $\rho=\rho (r,t)$ is the local matter density.  The condition for a trapping surface to be formed follows from the constraint that the gravittational radius is
\be
R_{S}(r,t)\geq r.
\label{trap}
\ee
for a given value of coordinates $(t,r)$.    If the source is completely contained within this region, then $R_S$ is identified with the usual Schwarzschild radius.  More generally, the condition (\ref{trap}) gives a more rigorous representation of the hoop conjecture \cite{Thorne:1972ji}, which allows for the formation of a black hole in the collision of two masses if their impact parameter $b$ is contained within the Schwarzschild radius.  From the above definitions, this can be re-expressed as the condition
\be
b\lesssim 2\lp E/\mpl\equiv\rh,
\label{impact}
\ee
where E is the total energy in the centre-of-mass frame. 

Since such an object would be manifestly quantum mechanical, one must also introduce an uncertainty in its position.  This will be on the order of the system's Compton wavelength, $\lambda_{m}\simeq\lp \mpl/m$, providing the additional constraint on the gravitational radius
\be
R_{S}\gtrsim \lambda_{m}\; \; \; \; \Longrightarrow \; \; \; \; m\gtrsim \mpl
\ee
This spread in localization can be represented by the wavefunction
\be
\ket{\psi_{S}}=\sum_{E}C(E)\ket{\psi_{E}} \;,
\ee
As usual, the sum over the variable $E$ represents the decomposition on the spectrum of the Hamiltonian,
\be
\hat{H}\ket{\psi_{E}}=E\ket{\psi_{E}}
\ee
Once the energy spectrum is known, we can use (\ref{impact}) to get 
\be
E=\mpl\frac{\rh}{2\lp}.
\label{schwarzE}
\ee
One can now define the HWF as
\be
\psi_{H}(\rh)=C(\mpl \,\rh/2\,\lp) \;,
\ee
which can be normalized as
\be
\langle\psi_{H}|\phi_{H}\rangle=4\pi \int_{0}^{\infty}\psi_{H}^{*}(\rh)\phi_{H}(\rh)\rh^{2}\d\rh .
\label{normalize}
\ee
Conceptually, the normalized HWF $\psi_{H}$ yields the probability for an observer to measure particle in the quantum state $\psi_{S}$ and associate to it a horizon of radius $r=r_{H}$. Consequently, the sharply defined classical radius is replaced by the expectation radius of the operator $\hat{r}_{H}$.

The probability for the source to be a black hole is that it lies completely within its horizon,
\be
{\cal {P}}_{BH}=\int_{o}^{\infty}{\cal {P}}_{<}(r <\rh) \d \rh.
\ee
where the density
\be
{\cal {P}}_{<}(r <\rh)={\cal {P}}_{S}(r<\rh){\cal {P}}_{H}(\rh) 
\ee
is a combination of requiring the particle to rest within a sphere of radius $r = r_H$, and the probability that $r_H$ is the gravitational radius.  These are respectively calculated as
\be
{\cal {P}}_{S}(r<\rh)=\int_{0}^{\rh}{\cal {P}}_{S}(r)\d r=4\pi \int_{0}^{\rh} |\psi_{S}(r)|^2 r^2 \d r
\ee
and
\be 
{\cal {P}}_{H}(\rh)=4\pi r^2_{H} |\psi_{H}(\rh)|^2
\ee

\section{Generalized Uncertainty Principle Black Holes}
\label{gupreview}
\setcounter{equation}{0}

As one approaches the Planck scale, it has been argued \cite{Veneziano:1986zf,Kempf:1994su,Adler_1} that the Heisenberg Uncertainty Principle (HUP) should be replaced by a Generalized Uncertainty Principle (GUP) of the form
\be
\Delta x >\frac{\hbar}{\Delta p}+\left(\frac{\alpha\lp^2}{\hbar}\right)\Delta p
\ee
where $\alpha$ is a dimensionless constant that depends on the particular model of interest.  This introduces a duality in the momentum uncertainty of the form $\Delta x \sim \Delta p + \frac{1}{\Delta p}$, and assuming the correspondnace $\Delta p \rightarrow m$, one can determine a similar mass duality to be present in the characteristic length scale of the system.

The basis of the argument in \cite{Carr:2015nqa} is that the GUP is encoded in the spacetime geometry, and so a metric description of spacetime must incorporate such a mass duality.
In the large mass limit $M\gg\Mpl$ where quantum effects are negligible, one should recover the Schwarzschild solution.  In this case, the black hole mass is well-defined in terms of the stress energy tensor.  When $M<\Mpl$, however, the exact meaning of the mass parameter becomes ambiguous, referring to both a particle and a black hole.
Since the horizon radius of sub-Planckian mass black holes would be shorter than the Planck length itself, the relativistic description becomes unreliable.  Consequently, one takes the object to be a particle of mass $M \sim \frac{\hbar}{\lambda_C}$, with $\lambda_C$ the Compton wavelength.  This can also be expressed as a variant of the Komar mass
\begin{equation}
M\equiv \int_{\Sigma} d^3x \sqrt{\gamma}\ n_\mu K_\nu T^{\mu\nu}\simeq -4\pi\int_0^{\lambda_C} dr \, r^2  T^{\ 0}_{ 0}
\label{komarparticle}
\end{equation}
where $\gamma$ is the determinant of the spatially induced metric $\gamma^{ij}$, $T^{\mu \nu}$ is the stress-energy tensor and $T^{\ 0}_{ 0}$ quantifies the energy density over the length scale $\lambda_C$.

Since we lack a full quantum theory of gravity, however, the exact form of the stress energy tensor above is nebulous, and so one can assume the value of $T^0_0$ to represent some quantum mechanical distribution of matter \cite{Carr:2015nqa}.  Consequently, the full definition of the mass will include both the large scale ({\it e.g.} ADM) mass, as well as the short scale particle mass.

Incorporating this new relation for the mass, one arrives at a quantum corrected form of the Schwarzschild metric \cite{Carr:2015nqa}
\be
ds^2=F(r)dt^2-F(r)^{-1}dr^2-r^2d\Omega^2 \\
F(r)=1-\frac{2}{M_{Pl}^2}\frac{M}{r}\left(1+\frac{\beta}{2}\frac{M_{Pl}^2}{M^2}\right)
\ee
In essence, this metric encapsulates all features of the Schwarzschild solution by virtue of the fact that the modification in the mass term is coordinate-independent.   Furthermore, natural dimensional reduction features are demonstrated in the gravitational radius and thermodynamics of Sub-Planckian objects ($m \ll \Mpl$) that resemble that of (1+1)-D gravity.

Specifically, the horizon is
\be
\rh=\frac{2}{M_{Pl}^2}\left(\frac{M^2+\frac{\beta}{2}M_{Pl}^2}{M}\right)
\label{guprh}
\ee
which yields 
\be
M\gg M_{Pl} \;\; \Longrightarrow \;\; \rh\approx\frac{2M}{M_{Pl}^2}\\
M\approx M_{Pl} \;\; \Longrightarrow \;\;\; \rh\approx\frac{2+\beta}{M_{Pl}} \\
M\ll M_{Pl} \;\; \Longrightarrow \;\; \rh\approx\frac{\beta}{M}
\ee
for super-Planckian, Planckian, and sub-Planckian mass black holes.

\begin{figure}[H]
\hspace*{5cm}\includegraphics[scale=0.7]{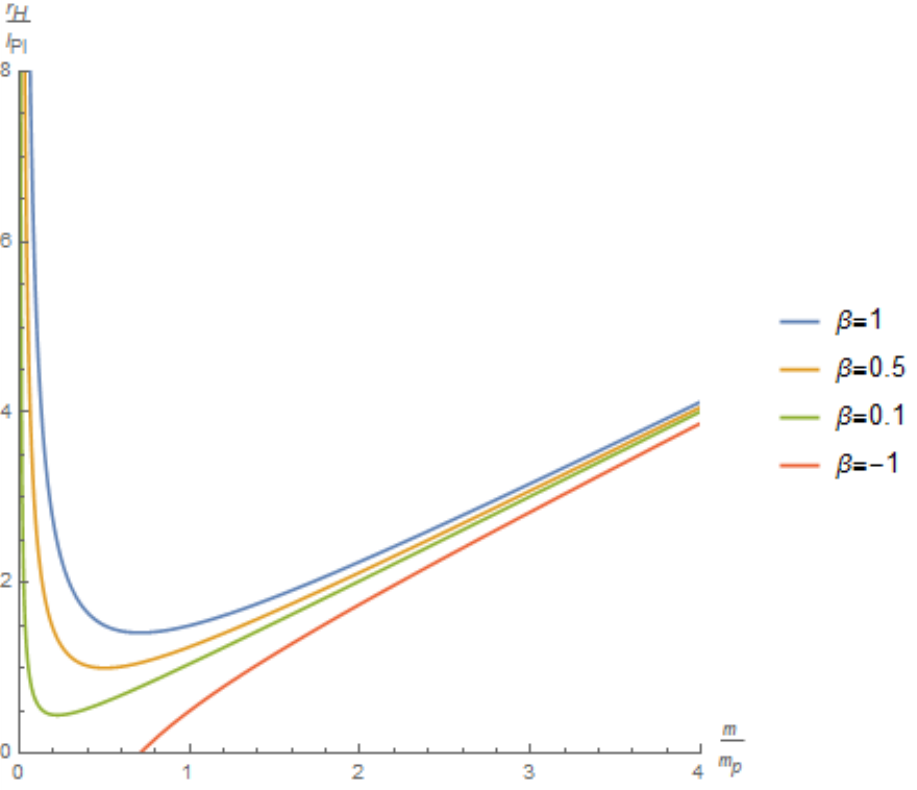}
\caption{Horizon radius (\ref{guprh}) as a function of mass $M/\mpl$ for $\beta=1$ (red, top), $\beta=0.5$ (dark blue, second from top), $\beta=0.1$ (green, third from top) and $\beta=-1$ (lighter blue, bottom). In the last case, the horizon vanishes when $M=\sqrt{|\beta|/2}M_{Pl}$ and is defined only for bigger masses than this cut-off.  This Figure is reproduced from \cite{Carr:2015nqa}.}
\label{fig0}
\end{figure}

In \cite{Carr:2015nqa} it is shown how although the singularity is not removed, it can never be reached. This is done by inverting (\ref{guprh}) in terms of the two masses associated to a given horizon radius $\rh$, which give a minimum radius 
\be
r_{min}=2\sqrt{2\beta}\lp,
\ee
and an associated mass
\be
M(r_{min})=\sqrt{\beta/2}M_{Pl}
\ee
Therefore, $\rh\geq r_{min}$ so the singularity can never be externally reached.

\section{HWF of GUP Black Holes}

As previously done in \cite{Casadio:2013tza,Casadio:2013aua,Casadio:2013uga,Casadio:2015jha,Casadio:2016fev}, we can describe a massive particle at rest in the origin of a reference frame with the spherically symmetric Gaussian wavefunction
\be
\psi_{\rm S}(r)=\frac{e^{-\frac{r^2}{2\,\ell^2}}}{(\ell\,\sqrt{\pi})^{3/2}}.
\ee

Inspired by the dual role of $m$ in the GUP, we now explore the existence of sub-Planckian black holes, i.e quantum mechanical objects that are simultaneously elementary particles and black holes. In this context, we replace our usual mass $m$ by the ``GUP'' mass $M$,
\be
M \equiv m\left( 1+\frac{\beta}{2}\frac{\mpl^{2}}{m^{2}}\right) 
\ee

Next, we consider the particular case where the width $\ell$, related to the uncertainty in the size of the particle, is approximately given by the Compton length. 
\be
\ell=\lambda_{m}\simeq \lp \frac{\mpl}{M}.
\ee

Noting that because the analysis holds for independent $\ell$ and $m$, such case corresponds to maximum localisation for the source as one expects $\ell \geqslant \lambda_{m}$.

Taking the Fourier Transform, the corresponding wavefunction in momentum space gives
\be
\tilde{\psi}_{\rm S}(p)=\frac{e^{-\frac{p^2}{2\,\Delta^2}}}{(\Delta\, \sqrt{\pi})^{3/2}}
\ee
where $\Delta=\mpl \,\lp/\ell$ is the spread of the wave-packet in momenta space.

Assuming the relativistic mass-shell equation in flat space-time to account for high-energy particle collisions, we relate the momentum $p$ to the total energy $E$ 
\be
E^{2}=p^{2}+M^{2}
\ee

From the relation for the Schwarzschild radius (\ref{impact}) and fixing the normalization by means of (\ref{normalize}), we then obtain the HWF
\be
\psi_{\rm H}(\rh)=\frac{1}{4\lp^{3}}\, \sqrt{\frac{\ell^{3}}{\pi \Gamma\left(\frac{3}{2},1\right)}}\, \Theta(\rh-R_{\rm H})\, e^{-\frac{\ell^{2}\rh^{2}}{8\ell^{4}}}
\ee
where we defined $R_{\rm H}=2\lp M/\mpl$ and the Heaviside step function arises from the fact that $E\geqslant M$. Finally, 
\be
\Gamma(s,x)=\int_{x}^{\infty} t^{s-1}e^{-t} dt\, ,
\ee
is the upper incomplete Gamma function.

One can calculate the probability density to be
\be
P_{<}=\frac{\ell^{3}}{2\sqrt{\pi}\lp^{6}}\frac{\gamma\left(\frac{3}{2},\frac{\rh^{2}}{\ell^{2}}\right)}{\Gamma\left(\frac{3}{2},1\right)}\, \Theta(\rh-R_{\rm H})\, e^{-\frac{\ell^{2}\rh^{2}}{4\ell^{4}}}\, \rh^{2}
\ee
where $\gamma(s,x)=\Gamma(s)-\Gamma(s,x)$ is the lower incomplete Gamma function.

Integrating the density for $\rh$ from $R_{\rm H}$ to infinity gives us the probability of a particle to be a black hole in terms of its mass
\begin{multline}
{\cal {P}}_{BH}(m)=-\frac{2 \sqrt{\pi } T\left(2 \sqrt{2} m^2 \left(\frac{\beta }{2 m^2}+1\right)^2,\frac{1}{2 m^2 \left(\frac{\beta }{2 m^2}+1\right)^2}\right)}{\Gamma
   \left(\frac{3}{2},1\right)}+\text{erf}\left(2 m^2 \left(\frac{\beta }{2 m^2}+1\right)^2\right)+ \\  \frac{\sqrt{\pi } \text{erfc}\left(2 m^2 \left(\frac{\beta }{2
   m^2}+1\right)^2\right)}{2 \Gamma \left(\frac{3}{2},1\right)}-\frac{2 m^2 e^{-4 m^4 \left(\frac{\beta }{2 m^2}+1\right)^4-1} \left(4 m^4 \left(\frac{\beta }{2
   m^2}+1\right)^4+3\right) \left(\frac{\beta }{2 m^2}+1\right)^2}{\sqrt{\pi } \Gamma \left(\frac{3}{2},1\right) \left(4 m^4 \left(\frac{\beta }{2 m^2}+1\right)^4+1\right)^2}
\end{multline}
where \;$T(h,a)=\frac{1}{2\pi}{\displaystyle \int_{0}^{a} }\frac{e^{-\frac{1}{2}h^2(1+x^2)}}{1+x^2}\d x$\; is Owen's function.
\bigskip

\begin{figure}[H]
\hspace*{3cm}\includegraphics[scale=0.4]{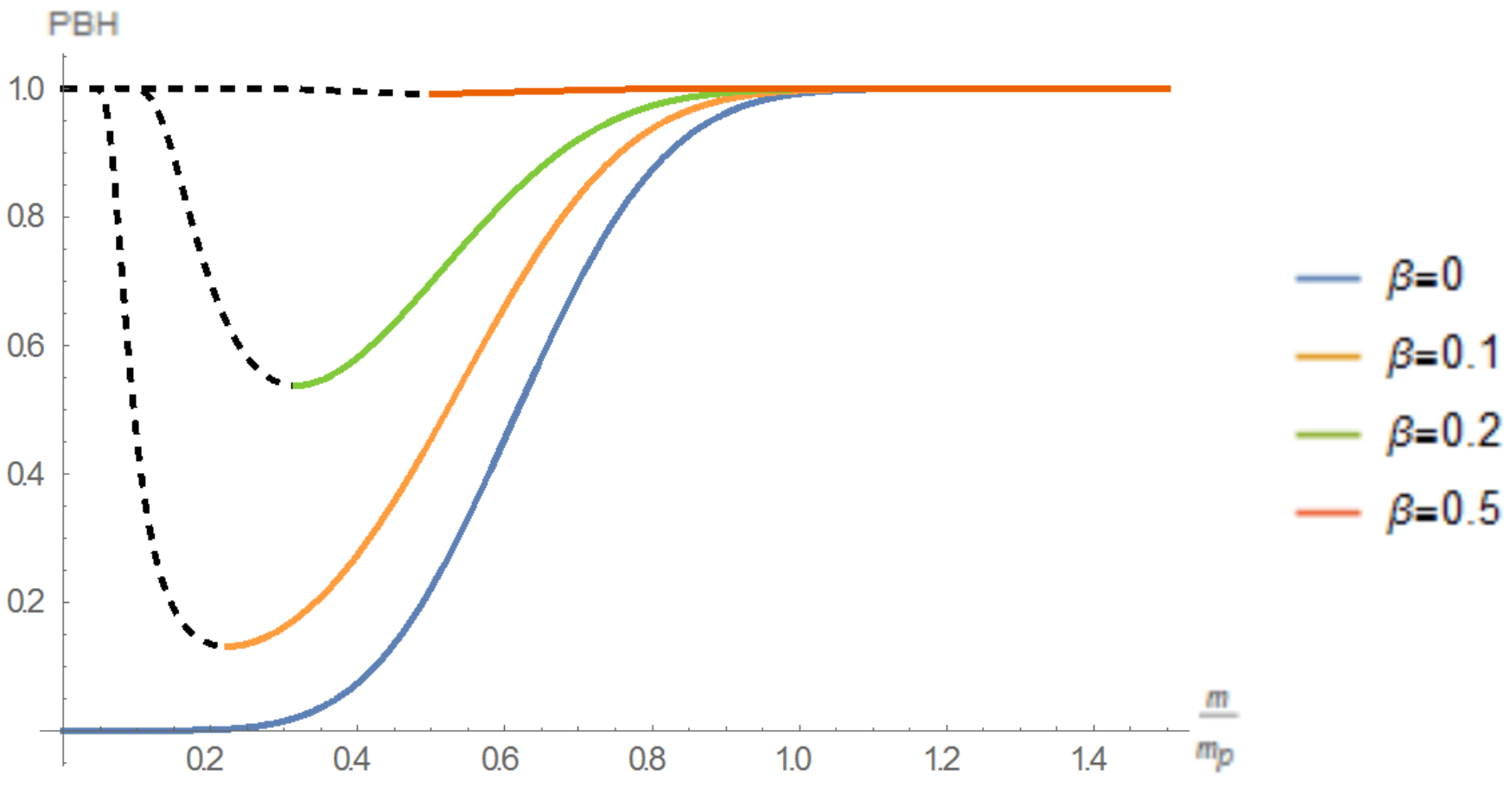}
\caption{Probability $P_{\rm BH}(m)$ for a particle to be
a black hole for increasing values of the free parameter $\beta$ vs. mass of the particle. Dashed lines correspond to values where $m/\mpl<\sqrt{\beta/2}M_{Pl}$.}
\label{fig1}
\end{figure}

The associated probabilities are shown in Figure~\ref{fig1} for positive $\beta$. We restrict this scenario to the cut-off mass shown before to correspond to the minimum radius $r_{min}$ given by $M(r_{min})=\sqrt{\beta/2}M_{Pl}$; and masses below this cut-off are evaluated for numerical purposes only. Nevertheless, we remind the reader that in the original GUP formulation \cite{Carr:2015nqa} there is no minimum mass restriction, therefore the limit $M \to 0$ can theoretically be reached.  First, we note that for vanishing $\beta$ the results for the probability of an elementary particle to be a black hole resemble that of the standard Schwarzschild metric \cite{Casadio:2015qaq}. This is in agreement with our modified theory, given that for $\beta=0$, $M \to m$ and the horizon radius becomes the classical Schwarzschild radius. 

On the other hand, for increasing $\beta$, the graph displays a minimum in the probability, meaning that for every value of the mass there is a certain chance that the particle will be a black hole. Furthermore, for sufficiently large $\beta$, everything is a black hole, in agreement with the fact the GUP imposes no minimum mass, and that the effect of increasing this free parameter translates to having a bigger mass confined in a bigger horizon radius, thus making a particle to be a black hole more probable. In analogy with the Heisenberg Uncertainty Principle (HUP), the dashed regime of the graphs for $m \to 0$ indicates that $\Delta p \to 0$ so $\Delta x \to \infty$ like the case of a free particle in Quantum Mechanics, hence ${\cal {P}}_{BH}\simeq 1$ and making it a certainty that the particle will lie somewhere within its horizon radius and thus \textit{be} a black hole. 
By taking the limit $R_{\rm H} \to 0$ in the Heaviside function, we obtain a simple analytic approximation for the HWF,
\be
\psi_{\rm H}(\rh)=\left(\frac{\ell}{2\sqrt{\pi}\lp^{2}}\right)^{3/2} e^{-\frac{\ell^{2}\rh^{2}}{8\ell^{4}}}
\ee
from which by the same procedure, one can derive the approximate probability in terms of the mass to be
\be
P_{BH}(m)=\frac{\frac{2 m \cot ^{-1}\left(\frac{2 m^2}{\left(\beta +2 m^2\right)^2}\right)}{\beta +2 m^2}-\frac{2 \left(\frac{8 m^3}{\left(\beta +2 m^2\right)^3}-\frac{2 \beta }{m}-4 m\right)}{\left(4 \beta +\frac{\beta ^2}{m^2}+4 m^2 \left(\frac{1}{\left(\beta +2 m^2\right)^2}+1\right)\right)^2}}{\pi  \sqrt{\frac{m^2}{\left(\beta +2 m^2\right)^2}}}
\ee

\begin{figure}[H]
\hspace*{3cm}\includegraphics[scale=0.4]{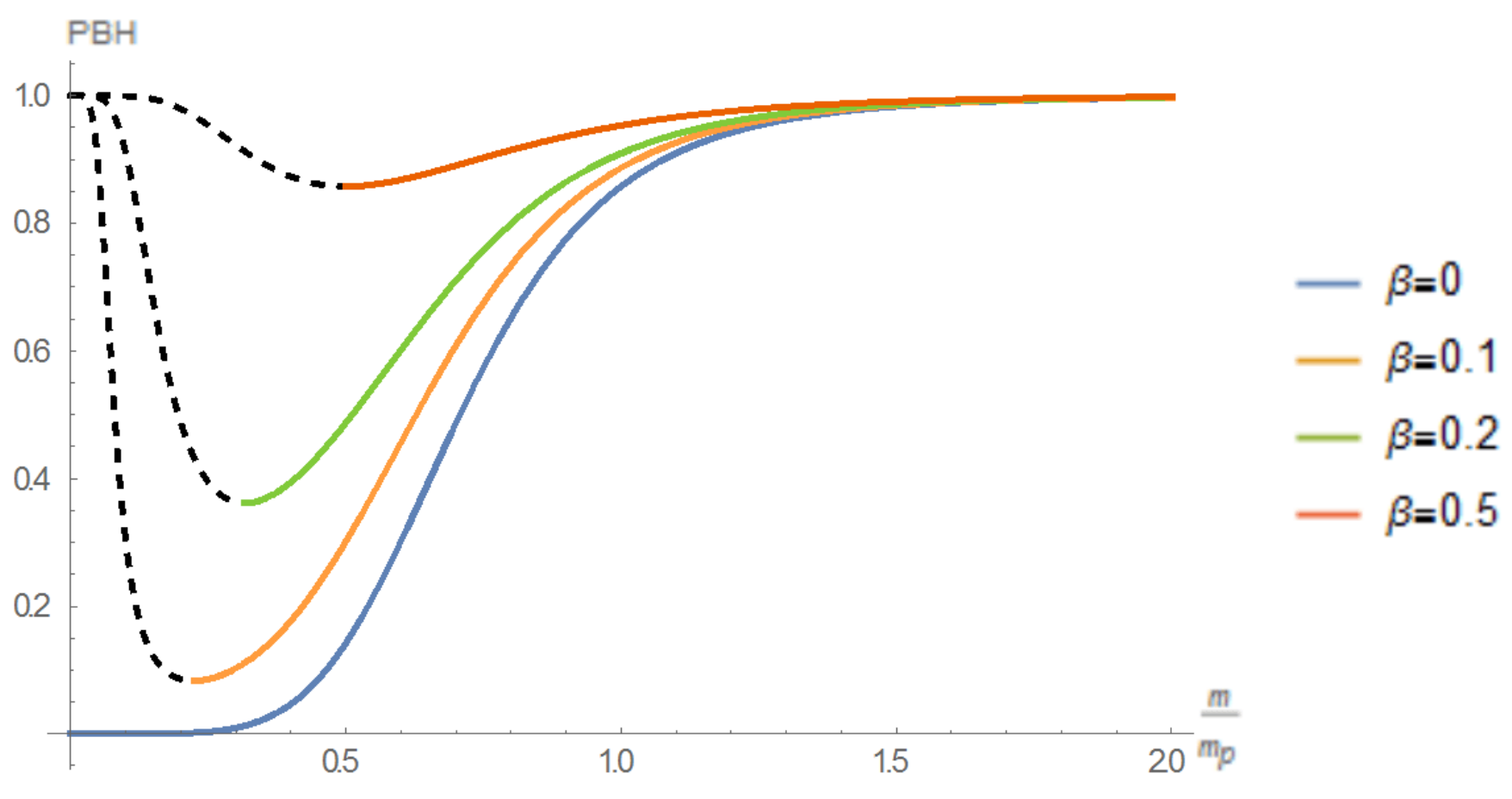}
\caption{Analytic approximation of the probability $P_{\rm BH}(m)$ for a particle to be a black hole vs. mass of the particle for increasing values of the free parameter $\beta$ .Dashed lines correspond to values where $m<\sqrt{\beta/2}M_{Pl}$.}
\label{fig2}
\end{figure}

Figure~\ref{fig2} shows graphically that this probability approximation is an slight underestimate of the probability in Eq. (3.19).

\begin{figure}[H]
\hspace*{3cm}\includegraphics[scale=0.8]{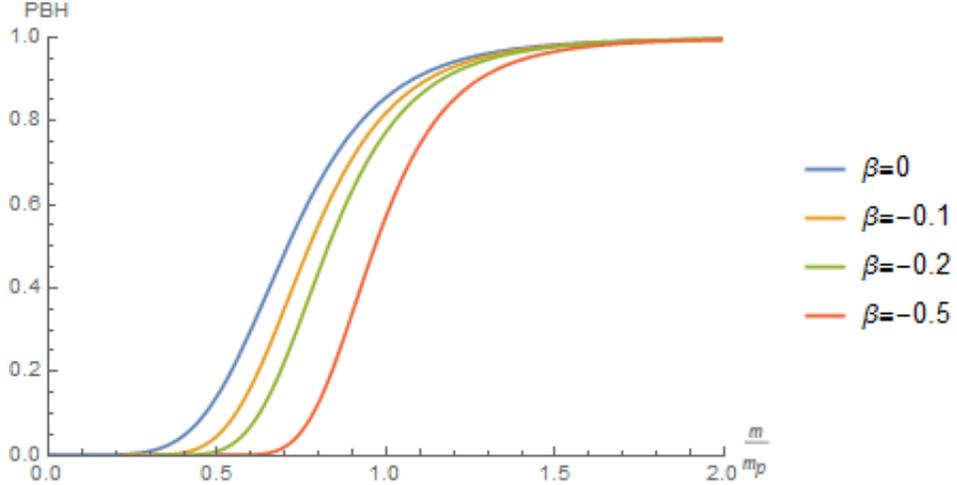}
\caption{Analytic approximation of the probability $P_{\rm BH}(m)$ for a particle to be a black hole vs. mass of the particle for decreasing values of the free parameter $\beta$ .}
\label{fig3}
\end{figure}

Next, negative values of $\beta$ are considered in Figure~\ref{fig3}. This case also presents a minimum allowed mass given by $M>\sqrt{|\beta|/2}M_{Pl}$. Nevertheless, by the time this cut-off comes to action ${\cal {P}}_{BH}\simeq 0$ so its effect goes unnoticed in the graphs. In this case, decreasing $\beta$ is a smooth extension of the Schwarzschild scenario that simply shifts the probability curve to the right, given that $M$ and $\rh$ both get smaller, thus making it more improbable for the particle to be a black hole.

An important feature in Figures~\ref{fig1},\ref{fig2},\ref{fig3} is that $P_{BH} \simeq 1$ for $m \gtrsim \mpl$.

Similarly, the probability can be expressed in terms of the Gaussian width assuming that $\ell/\lp=\mpl/m$.
\be
P_{BH}(\ell)=\frac{|\ell \beta +\frac{2}{\ell}|\left(\frac{2 \ell \cot^{-1}\left(\frac{2 \ell^2}{\left(\beta  \ell^2+2\right)^2}\right)}{\beta  \ell^2+2}-\frac{2 \left(\frac{8 \ell^3}{\left(\beta  \ell^2+2\right)^3}-2 \beta  \ell-\frac{4}{\ell}\right)}{\left(4 \beta +\ell^2 \left(\beta ^2+\frac{4}{\left(\beta \ell^2+2\right)^2}\right)+\frac{4}{\ell^2}\right)^2}\right)}{\pi}
\ee

\begin{figure}[H]
\hspace*{3cm}\includegraphics[scale=0.4]{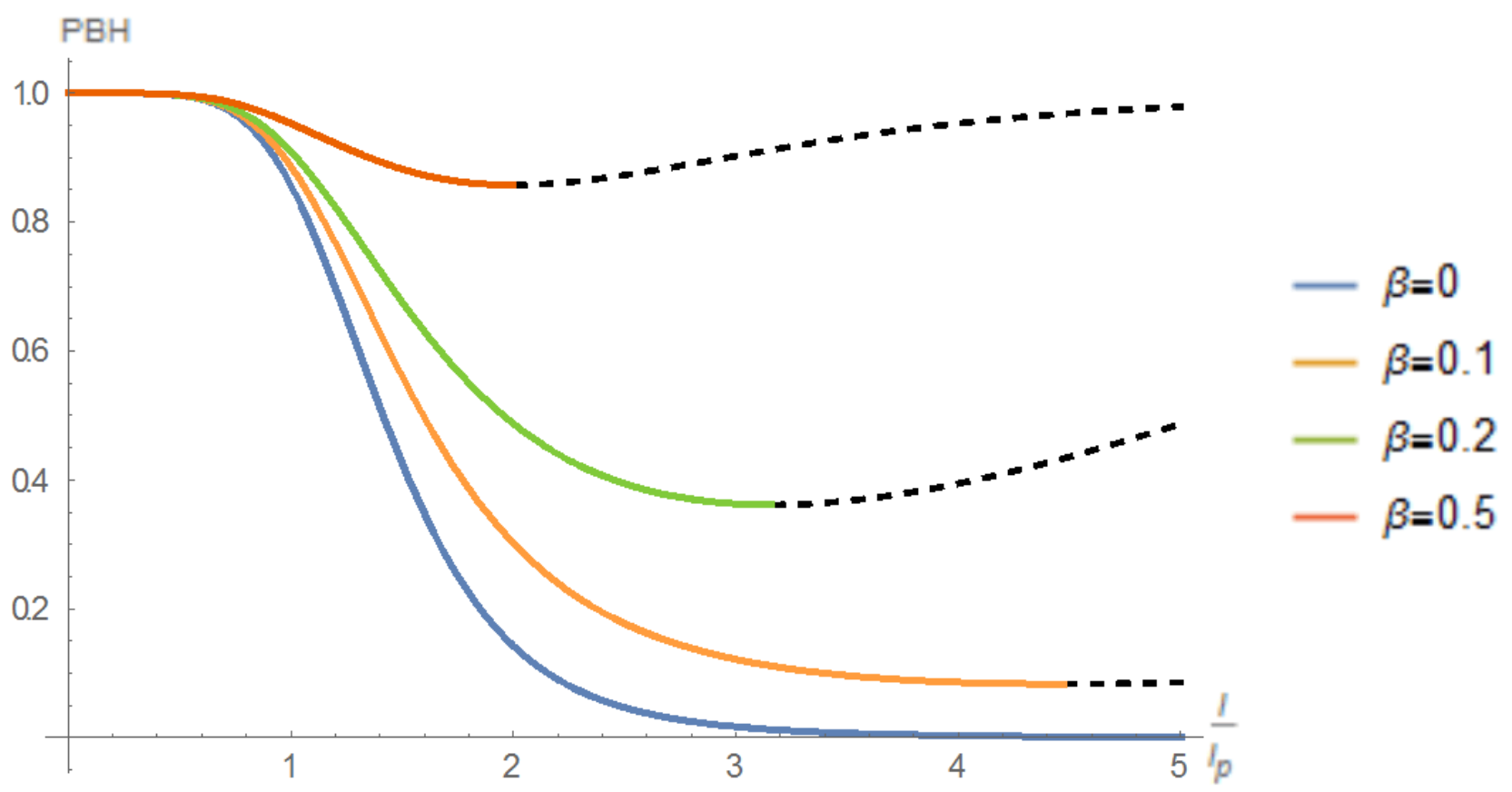}
\caption{Analytic approximation of the probability $P_{\rm BH}(m)$ for a particle to be a black hole vs. Gaussian width for increasing values of the free parameter $\beta$. Dashed lines correspond to the values $\ell>\sqrt{2/\beta}\;\lp$.}
\label{fig4}
\end{figure}

In Figure~\ref{fig4}, one can see that increasing $\beta$ induces a probability for the particle to be a black hole for larger Gaussian width values, which corresponds to a big localisation, in agreement with the results of Figure~\ref{fig2}. Furthermore, the minimum mass cut-off \linebreak $M>\sqrt{|\beta|/2}M_{Pl}$ induces a maximum Gaussian width cut-off given by 
\be
\ell>\sqrt{\frac{2}{\beta}}\lp
\ee
so no particle can physically be localized in a width bigger than the above. This may be an indication of some phenomena occurring around this scale, which can correspond to a phase transition of the black hole into something else. 

\begin{figure}[H]
\hspace*{3cm}\includegraphics[scale=0.8]{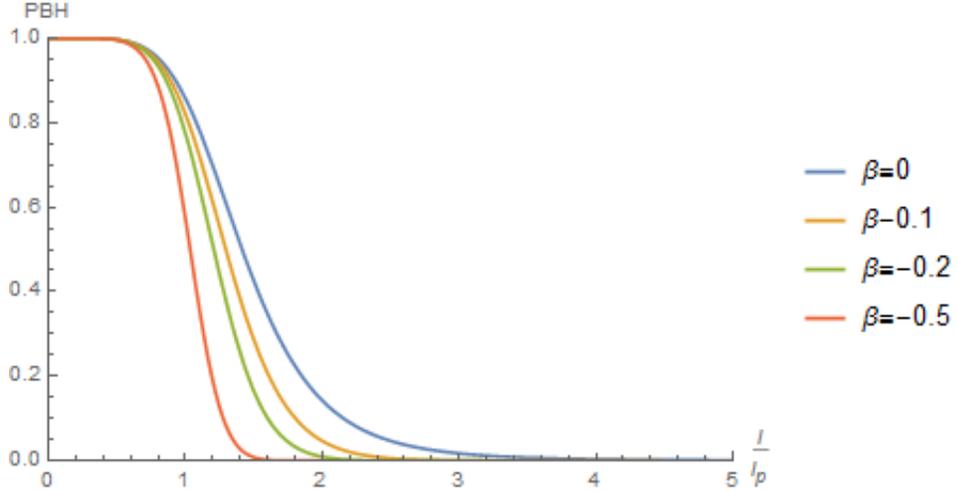}
\caption{Analytic approximation of the probability $P_{\rm BH}(m)$ for a particle to be a black hole vs. Gaussian width for decreasing values of the free parameter $\beta$ .}
\label{fig5}
\end{figure}

Finally, Figure~\ref{fig5} illustrates how decreasing $\beta$ causes the probability of the particle to be a black hole diminish for a given Gaussian width value. Again, the negative $\beta$ scenario ends up being unaffected by the imposed cut-off as before, because ${\cal {P}}_{BH}\simeq 0$ when the maximum allowed localization is reached. 

It is important to emphasize in Figures~\ref{fig4},\ref{fig5} that for $\ell \lesssim \lp \to P_{BH} \simeq 1$ independent of $\beta$, which is a desired intuitive result.

\section{Conclusions}

In this paper we extended the results of \cite{Casadio:2013tma} by embedding the source in a GUP inspired metric discussed on \cite{Carr:2015nqa}. Considering the case of a wave-packet shaped by a Gaussian distribution, the corresponding horizon wavefunction was computed and the probability ${\cal {P}}_{BH}$ that the source is a (quantum) black hole, i.e that it lies within its horizon radius, was calculated. 

The case for $\beta<0$ looks qualitatively similar to the standard Schwarzschild case, which is recovered when $\beta=0$. The general shape of ${\cal {P}}_{BH}$ is maintained when decreasing the free parameter, and shifted to reduce the probability for the particle to be a black hole accordingly. Effectively, a negative $\beta$ reduces the magnitude of $M$ and consequently $R_{H}$, thus matching our intuition that either the particle should be more localized or more massive in order to be a black hole.	

The $\beta>0$ scenario differs in many ways from the previous one. First of all, when increasing the free parameter a minimum in ${\cal {P}}_{BH}$ is encountered, thus meaning that every particle has some probability of decaying to a black hole. Furthermore, for sufficiently large $\beta$ one gets that every particle is a quantum black hole. Again, this results are in agreement with the intuitive effect of increasing $\beta$, which creates larger $M$ and $R_{H}$ terms, thus making it more probable for the particle to lie within in horizon radius and hence be a black hole.

To conclude, although in our approach we restrict the sub-Planckian regime by imposing a mass/localisation cut-off, the exact nature of extremely low mass black holes ultimately depends on whatever is the correct model of quantum gravity. It might be the case that this is something described by \textit{e.g.} the quantum $N$ portrait approach for low graviton number \cite{Dvali:2015rea,Dvali:2012wq,Frassino:2016oom}.

Finally, possible ways of extending this investigation include introducing the GUP formulation of HQM by modifying the momentum-space wavefunction with one that encodes the generalized uncertainty relation
\be
\psi_{\xi}^{ml}(p)=\sqrt{\frac{2\sqrt{\beta}}{\pi}}(1+\beta p^2)^{-\frac{1}{2}} e^{-i\frac{\xi tan^{-1}(\sqrt{\beta}p)}{\hbar\sqrt{\beta}}}
\ee 
as proposed in \cite{Kempf:1994su}. Furthermore, future research can analyze the effect of modifying the Heaviside function in (3.16), to one where the edges are smoothed by the ``golden rule'' \cite{Mureika:2011hg} because of the uncertainty introduced at the quantum scale. 

\vskip 1cm
\noindent {\bf Acknowledgments}\\
LM and JM thank Roberto Casadio and Piero Nicolini for insightful discussions on the manuscript.  LM was supported by a Frank R.\ Seaver Summer Undergraduate Research Scholarship from Loyola Marymount University.

\end{document}